\documentclass[%
reprint,
superscriptaddress,
 amsmath,amssymb,
 aps,
]{revtex4-1}

\usepackage[dvipdfmx]{color}

\usepackage{graphicx}
\usepackage{dcolumn}
\usepackage{bm}
\usepackage{physics}%

\usepackage{hyperref}

\newcommand{\red}[1]{\textcolor{red}{#1}}

\newcommand{\casea}{\text{(\red{A})}}
\newcommand{\caseb}{\text{(\red{B})}}

\usepackage{braket}
\usepackage{mathtools}
\usepackage{comment}

\begin{document}
\preprint{-----------}

\title{
Symmetry-protected quantization of complex Berry phases in non-Hermitian many-body systems
}

\author{Shoichi Tsubota}
\affiliation{%
Department of Physics, 
Graduate School of Science,
The University of Tokyo, 7-3-1 Hongo, Bunkyo-ku, Tokyo 113-0033, Japan}%

\author{Hong Yang}
\affiliation{%
Department of Physics, 
Graduate School of Science,
The University of Tokyo, 7-3-1 Hongo, Bunkyo-ku, Tokyo 113-0033, Japan}%

\author{Yutaka Akagi}
\affiliation{%
Department of Physics, 
Graduate School of Science,
The University of Tokyo, 7-3-1 Hongo, Bunkyo-ku, Tokyo 113-0033, Japan}%

\author{Hosho Katsura}
\affiliation{%
Department of Physics, 
Graduate School of Science,
The University of Tokyo, 7-3-1 Hongo, Bunkyo-ku, Tokyo 113-0033, Japan}%
\affiliation{%
Institute for Physics of Intelligence, The University of Tokyo, 7-3-1 Hongo, Bunkyo-ku, Tokyo 113-0033, Japan}%
\affiliation{%
Trans-scale Quantum Science Institute, University of Tokyo, Bunkyo-ku, Tokyo 113-0033, Japan}%

\date{\today}
\begin{abstract}

We investigate the quantization of the complex-valued Berry phases in non-Hermitian quantum systems with certain generalized symmetries. In Hermitian quantum systems, the real-valued Berry phase is known to be quantized in the presence of certain symmetries, and this quantized Berry phase can be regarded as a topological order parameter for gapped quantum systems. In this paper, on the other hand, we establish that the complex Berry phase is also quantized in the systems described by a family of non-Hermitian Hamiltonians. Let $H(\theta)$ be a non-Hermitian Hamiltonian parameterized by $\theta$. Suppose that there exists a unitary and Hermitian operator $P$ such that $PH(\theta)P = H(-\theta)$ or $PH(\theta)P = H^\dagger(-\theta)$. We prove that in the former case, the complex Berry phase $\gamma$ is $\mathbb{Z}_2$-quantized, while in the latter, only the real part of $\gamma$ is $\mathbb{Z}_2$-quantized. The operator $P$ can be viewed as a generalized symmetry operation for $H(\theta)$, and in practice, $P$ can be, for example, a spatial inversion.
Our results are quite general and apply to both interacting and non-interacting systems. We also argue that the quantized complex Berry phase is capable of classifying non-Hermitian topological phases, and demonstrate this in one-dimensional many-body systems with and without interactions.
\end{abstract}

\maketitle

\textit{Introduction. ---}
The study of topological phases is one of the major themes in modern condensed matter physics \cite{hasan2010,qi2011,senthil2015,chiu2016,wen2017} since the discovery of the integer quantum Hall effect (IQHE)~\cite{klitzing1980}. Topological phases are known to be characterized by topological invariants, which reflect the geometric properties of the wave functions and are stable 
under continuous deformations. For instance, IQHE is characterized by the Chern number~\cite{thouless1982, kohmoto1985}. However, characterizing the phases of 
strongly correlated systems in terms of topological invariants still remains challenging~\cite{fidkowski2010,shapourian2017,shiozaki2017}.
For systems with a finite excitation gap above the ground state, the Berry phase as a topological order parameter~\cite{hatsugai2006, hatsugai2007, hirano2008, hatsugai2011, motoyama2013, kariyado2014, chepiga2016, kariyado2018, mizoguchi2019Higher} is known to be one of the effective methods for analyzing topological phases. This method is based on the fact that the Berry phase~\cite{berry1984,zak1989,xiao2010} 
is quantized in the presence of certain symmetries such as time-reversal symmetry.
Physically, when the Berry phase defined by a local gauge twist on a bond is non-trivially quantized, it implies the existence of an entangled cluster over the bond.  
This quantized Berry phase, which is also applicable to interacting systems, can be 
used to effectively identify topological phase transitions.

In recent years, due to their relevance to open quantum systems, topological phases of systems described by non-Hermitian Hamiltonians~\cite{moiseyev2011,el2018,ashida2020} have attracted much attention. 
For examples, the Bulk-edge correspondence~\cite{Revision_Esaki2011,Revision_Lee2016,Revision_Xu2017,Revision_Leykam2017,Revision_Shen2018,Revision_Kunst2018,Revision_Yao2018,Revision_Jin2019,Revision_Yokomizo2019} and
the classification of topological phases~\cite{gong2018,kawabata2019,liu2019,xi2021} were intensively studied. 
However, characterization of topological phases in non-Hermitian systems is still difficult in general because the topological invariants defined for Hermitian systems may not directly apply to non-Hermitian ones \cite{yoshida2019non}. 
Thus, there is  
a need for methods that can be used to characterize topological phases in a wide variety of non-Hermitian systems.
One of such candidates is a complex Berry phase~\cite{garrison1988,dattoli1990,miniatura1990}, which is a generalization of the real-valued Berry phase to non-Hermitian systems. 
The complex Berry phase has proved useful in describing non-Hermitian topological phases in several models~\cite{Revision_Gong2010,liang2013,wagner2017,dangel2018,Revision_Zhang2019,Revision_Ezawa2021}
However, previous studies overwhelmingly focused on non-interacting systems.

In this paper, 
we prove that the complex Berry phase exhibits $\mathbb{Z}_2$ quantization in the presence of generalized symmetries described by unitary and Hermitian operators. 
Specifically, we discuss two classes of non-Hermitian Hamiltonians with such symmetries. 
In one class, the complex Berry phase $\gamma$ is quantized to $0$ or $\pi$. In the other class, $\gamma$ is not necessarily real, but its 
real part is shown to be quantized. 
While this quantization of the real part has been predicted in the case of non-interacting systems~\cite{liang2013,wagner2017,dangel2018,Revision_Ezawa2021}, we prove it in a more general setting, including interactions. 
This flexibility allows one to use the complex Berry phase as a topological order parameter to detect topological phases in interacting non-Hermitian systems. 
We demonstrate the power of this approach by numerically studying interacting fermion and spin models in one dimension. 

\medskip


\textit{Definition of the complex Berry phase. ---}
Let us introduce the definition of the complex Berry phase in the cyclic evolution in parameter space. Consider the parameter-dependent non-Hermitian Hamiltonian $H(\theta)$, 
where $\theta$ is a parameter taking values in $(-\pi, \pi]$,  corresponding to a closed loop in the parameter space. 
We identify $\theta=-\pi$ with $\pi$ so that $H(-\pi)=H(\pi)$. Note that generally $H(\theta)\neq H^\dag(\theta)$ and can include many-body interactions. 
In what follows, we assume that $H(\theta)$ is a smooth function of $\theta$ and that there is no exceptional point on the   
loop, i.e., $H(\theta)$ is diagonalizable for all $\theta$. Let $\ket{\Psi_n(\theta)} \ (\bra{\Phi_n(\theta)})$ be the $n$th right (left) 
eigenstate with 
eigenenergy $E_n(\theta) \in \mathbb{C}$. Note that one can always take the set of all right and left eigenvectors as a biorthogonal system such that $\braket{\Phi_n(\theta)|\Psi_m(\theta)}=\delta_{nm}$ for all $\theta$. 
In this paper, we call the state with the smallest real part of the energy the {\it ground state} 
and denote the corresponding right (left) eigenstate as $\ket{\Psi(\theta)}$  $(\bra{\Phi(\theta)})$. 
Further, we assume that the ground state is line-gapped \cite{kawabata2019} and unique
\footnote{This condition can be relaxed to the condition that $\ket{\Psi(\theta)} \ (\bra{\Phi(\theta)})$ is an isolated eigenstate such that no level crossing of the real part of energy occurs when the parameter $\theta$ is varied. 
}.
In other words, the unique ground state is separated from the other energy levels by a one-dimensional curve in the complex plane for all $\theta$.
Then, the ground state after the cyclic evolution is equal to the initial ground state multiplied by a phase factor. 
The complex Berry phase $\gamma$ acquired during this evolution is defined as 
\begin{align}
\gamma = -i\int_{-\pi}^{\pi} d\theta A(\theta) = -i\int_{-\pi}^{\pi} d\theta \braket{\Phi(\theta)|\partial_\theta|\Psi(\theta)},
\end{align}
where $A(\theta)=\braket{\Phi(\theta)|\partial_\theta|\Psi(\theta)}$ is the complex Berry connection for $\ket{\Psi(\theta)}$ and $\bra{\Phi(\theta)}$. 
The Berry phase is well-defined unless the energy gap closes in both finite and infinite systems \footnote{In the thermodynamic limit, an energy gap open in finite systems may close, and in such cases, the complex Berry phase is ill-defined.}.

\medskip

\textit{$\mathbb{Z}_2$ quantization of complex Berry phases. ---}
Let us suppose that there exists a unitary Hermitian operator $P=P^\dag = P^{-1}$ that satisfies either of the following two conditions:
\begin{align}
  \casea~ &
PH(\theta)P = H(-\theta), \quad ^\forall \theta \in (-\pi, \pi],
\label{inv1}
\\
  \caseb~ &
PH(\theta)P = H^{\dag}(-\theta), \quad ^\forall \theta \in (-\pi, \pi].
\label{inv2} 
\end{align}
Here, we assume that $P$ is independent of $\theta$. 
When we think of $\theta$ as a $U(1)$ twist on a link of a lattice model, we can take $P$ as a bond-centered inversion operator.
Note that in Hermitian systems,
Cases \casea \ and \caseb \ are identical. 
Under the above conditions, the complex Berry phase $\gamma$ for the unique and line-gapped ground state of the Hamiltonian $H(\theta)$ is $\mathbb{Z}_2$-quantized as 
\begin{align}
\casea~ &
\gamma 
\equiv 0, \pi \:\: ({\rm mod}\: 2\pi),
\\
\caseb~ &
{\rm Re}[\gamma]
\equiv 0, \pi \:\: ({\rm mod}\: 2\pi). 
\end{align}

We now prove these quantizations. 
Let $E(\theta)$ be the eigenenergy of the ground state $\ket{\Psi(\theta)}$ ($\bra{\Phi(\theta)}$). 
We can derive the relation between $E(\theta)$ and $E(-\theta)$ from the properties of similarity transformations. 
From Eq.~(\ref{inv1}), we have $E(\theta)=E(-\theta)$ in Case~\casea, while from Eq.~(\ref{inv2}), we have $E(\theta)=E^*(-\theta)$ in Case~\caseb. 
Then, from the uniqueness of the ground state and the
relation $\braket{\Phi(\theta)|P^2|\Psi(\theta)}=\braket{\Phi(\theta)|\Psi(\theta)}=1$ for all $\theta$,
one finds that $P$ operates on the right and left ground states as
\begin{align}
\casea~ 
P\ket{\Psi(\theta)} &=e^{i\chi(\theta)} \ket{\Psi(-\theta)},
\\
\bra{\Phi(\theta)}P &=e^{-i\chi(\theta)}\bra{\Phi(-\theta)},
\\
\caseb ~
P\ket{\Psi(\theta)} &=e^{i\zeta(\theta)} \ket{\Phi(-\theta)}=e^{i\zeta(\theta)} \bra{\Phi(-\theta)}^\dag, 
\\
\bra{\Phi(\theta)}P &=e^{-i\zeta(\theta)}\bra{\Psi(-\theta)}=e^{-i\zeta(\theta)}\ket{\Psi(-\theta)}^\dag,
\end{align}
\noindent
where $\chi(\theta)$ and $\zeta(\theta)$ are complex-valued phases depending on the parameter $\theta$. 
Next, by acting with $P$ on $\ket{\Psi(\theta)}$ twice, we have:
\begin{align}
\casea~ &
e^{i\chi(\theta)+i\chi(-\theta)}=1,
\label{twice1}
\\
\caseb~ &
e^{i\zeta(\theta)+i\zeta^*(-\theta)}=1.
\label{twice2} 
\end{align}
From Eq.~(\ref{twice1}), we see that $\chi(\theta)+\chi(-\theta) \equiv 0\: ({\rm mod}\: 2\pi)$ in Case~\casea, which, together with  the $P$ invariance at $\theta=0$ and $\pi$, implies that $\chi(0)$, $\chi(\pi) \equiv 0$, $\pi$ (${\rm mod}\: 2\pi$). 
Similarly, in Case~\caseb, Eq.~(\ref{twice2}) yields $\zeta(\theta)+\zeta^*(-\theta) \equiv 0\: ({\rm mod}\: 2\pi)$, implying that ${\rm Re}[\zeta(0)]$, ${\rm Re}[\zeta(\pi)] \equiv 0$, $\pi$ (${\rm mod}\: 2\pi$). 
We are now ready to prove the results. The relation between $A(\theta)$ and $A(-\theta)$ in each case is obtained as
\begin{align}
  \casea~ &
A(\theta)=\braket{\Phi(\theta)|P^2\partial_\theta|\Psi(\theta)}
=i{d\chi(\theta) \over d\theta} - A(-\theta),
\label{conn1}
\\
  \caseb~ &
A(\theta)=\braket{\Phi(\theta)|P^2\partial_\theta|\Psi(\theta)}
=i{d\zeta(\theta) \over d\theta} + A^*(-\theta).
\label{conn2} 
\end{align}
Then, by integrating Eqs.~(\ref{conn1}) and (\ref{conn2}) over $\theta$, we find that the complex Berry phases are $\mathbb{Z}_2$-quantized as
\begin{align}
  \casea~ &
\gamma 
= \int_{0}^{\pi}d\theta {d\chi(\theta) \over d\theta}
= \chi(\pi) - \chi(0) \equiv 0, \pi \:\: ({\rm mod}\: 2\pi),
\\
  \caseb~ &
{\rm Re}[\gamma]
={\rm Re}\left[\int_{0}^{\pi}d\theta {d\zeta(\theta) \over d\theta} \right]
\nonumber
\\
  & 
\:\:\:\:\:\:\:\:\:\:
= {\rm Re}[\zeta(\pi)]-{\rm Re}[\zeta(0)] \equiv 0, \pi \:\: ({\rm mod}\: 2\pi).
\end{align}
Note that in Case~\casea, the whole part of the complex Berry phase is guaranteed to be quantized, while in Case~\caseb, only the real part is quantized. In both cases, we can expect 
that the $\mathbb{Z}_2$ quantization is robust against small perturbations that satisfy either of the conditions Eqs.~(\ref{inv1}) and (\ref{inv2}).  
We note in passing that if $\theta$ is interpreted as a wavenumber, the quantized complex Berry phase can be considered as an extension of the Zak phase \cite{zak1989}.

Some remarks are in order. First, the proof for Case~\casea \ is essentially the same as that of the $\mathbb{Z}_2$-quantized Berry phase in a certain class of open quantum systems. See Ref.~\cite{yoshida2020} for more details. Second, in practical calculations, it is more convenient to rewrite the above formulas as
\begin{align}
\casea ~ &
 e^{i\gamma}=\braket{\Phi(\pi)|P|\Psi(\pi)}\braket{\Phi(0)|P|\Psi(0)},
\\
\caseb ~ &
e^{i{\Re}[\gamma]}=
{
\braket{\Psi(\pi)|P|\Psi(\pi)}
\over 
|\braket{\Psi(\pi)|P|\Psi(\pi)}|
}
{
\braket{\Psi(0)|P|\Psi(0)}
\over
|\braket{\Psi(0)|P|\Psi(0)}|
},
\end{align}
since they do not require the integration of the Berry connection over the path. 

In the following part, we numerically demonstrate the $\mathbb{Z}_2$-quantization of complex Berry phases in fermionic and spin models on finite lattices. 
We introduce the parameter-dependent Hamiltonian $H(\theta)$ by replacing a local term for a particular bond in $H(0)$ with a gauge-twisted one depending on $\theta$. As in the Hermitian case, the quantization of the Berry phase defined through this gauge twist allows one to detect entangled clusters such as spin singlets in the original Hamiltonian $H(0)$.
In other words, using the quantized complex Berry phase, we can detect topological phases characterized by such entangled objects. We show that this mechanism works well in our models.

\medskip

\begin{figure}[h]
\begin{center}
\includegraphics[width=7.8cm]{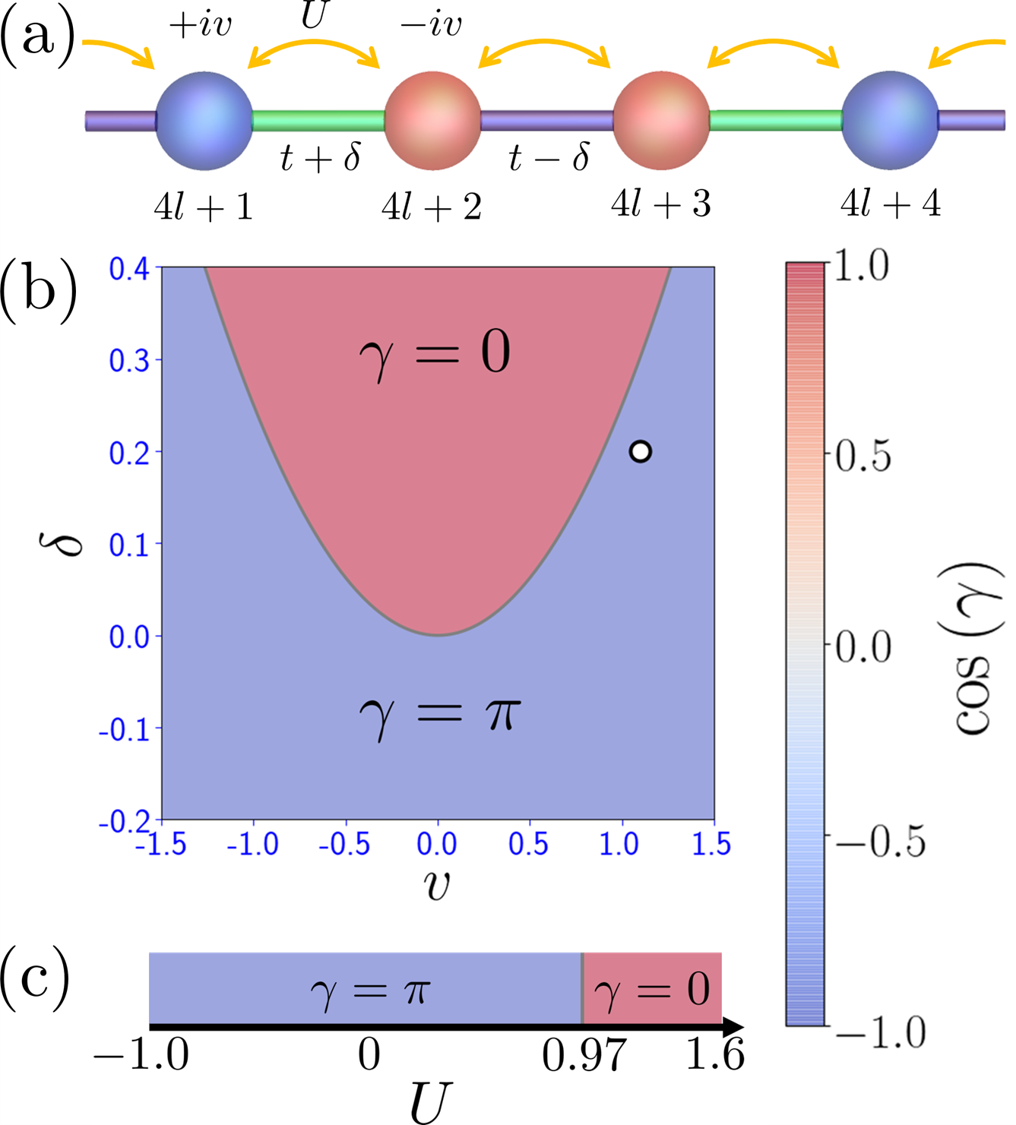}
\end{center}
\caption{(a) Schematic representation of 
the interacting fermion model Eq.~(\ref{Ham1}). (b) The complex Berry phase of the model Eq.~(\ref{Twist1}) 
for $t=1.0$, $U=0.0$. The system size is $N=64$ and the number of fermions is set to $N/2$ (half-filling). 
In the blue (red) region, 
$\gamma=\pi$ ($\gamma=0$). 
The gray solid line represents the phase boundary. 
(c) The complex Berry phase 
for Eq.~(\ref{Twist1}) with $(t, v, \delta)=(1.0, 1.1, 0.2)$. The system size is $N=16$ and the system is at half-filling. 
The corresponding parameter set is indicated by the white circle $\circ$ in~(b). 
}
\label{BP1}
\end{figure}

\textit{Non-Hermitian interacting fermion model. ---} 
As a demonstration for Case~\casea, 
we consider a system of interacting spinless fermions on a chain of length $N$ ($\in 4\mathbb{N}$) with periodic boundary conditions (PBC). We denote by $c^\dag_j$ $(c_j)$ the creation (annihilation) operator of a fermion at site $j$. As usual, the number operator is defined as $n_j=c^\dag_jc_j$. 
The system is described by the Hamiltonian
\begin{align}
H_1&=H_{\rm hop}+H_{\rm int}+H_{\rm pot},
\label{Ham1}
\end{align}
where
\begin{align}
H_{\rm hop} &= \sum_{j=1}^{N} \left\{ t - (-1)^j \delta \right\} \left(c_{j}^\dag c_{j+1} + c_{j+1}^\dag c_{j} \right),
\\
H_{\rm int}&= U\sum_{j=1}^{N} \left(n_j - {1 \over 2}\right)\left(n_{j+1} - {1 \over 2}\right),
\\
H_{\rm pot}&= iv \sum_{j=1}^{N}  \sqrt{2} \cos\left({\pi \over 2}j - {\pi \over 4}\right)n_{j}.
\end{align}
The first term $H_{\rm hop}$ is the Su-Schrieffer-Heeger
model~\cite{su1979,asboth2016}, 
where $t \pm \delta \in \mathbb{R}$ are staggered hopping amplitudes. The second term $H_{\rm int}$ describes the nearest-neighbor interaction with strength $U \in \mathbb{R}$. The last term $H_{\rm pot}$ represents the four-site periodic purely imaginary on-site potential with strength $v \in \mathbb{R}$. (See Fig.~\ref{BP1}(a) for a schematic picture.) 
We note in passing that the model is a generalization of the previously studied model~\cite{takata2018,brzezicki2019,comaron2020,Revision_Wu2021,Revision_Hyart2022}, which was designed to be realized in quantum dot arrays~\cite{Revision_Hyart2022}. In the following, we assume $t>0$ and $|\delta|<t$.

To define the complex Berry phase, we consider the model obtained by replacing the hopping between sites $N$ ($\in 4\mathbb{N}$) and $1$ with a complex one, depending on $\theta$. The resulting Hamiltonian reads
\begin{align}
H_1(\theta) = 
H_1 
+(t-\delta)\left\{(e^{-i\theta}-1)c^\dag_{N} c_{1}
+(e^{i\theta}-1)c^\dag_{1} c_{N}\right\}.
\label{Twist1}
\end{align}
In such a case, the Berry phase is also related to the polarization \cite{Revision_Resta1994}.
This Hamiltonian $H_1(\theta)$ satisfies Eq.~(\ref{inv1}) with the bond-centered inversion $P$: $c_j \rightarrow c_{N+1-j}$. 
Below, we focus on the complex Berry phase of the ground state at half-filling, i.e., the total fermion number is $N/2$. 

\begin{figure}[h]
\begin{center}
\includegraphics[width=8.2cm]{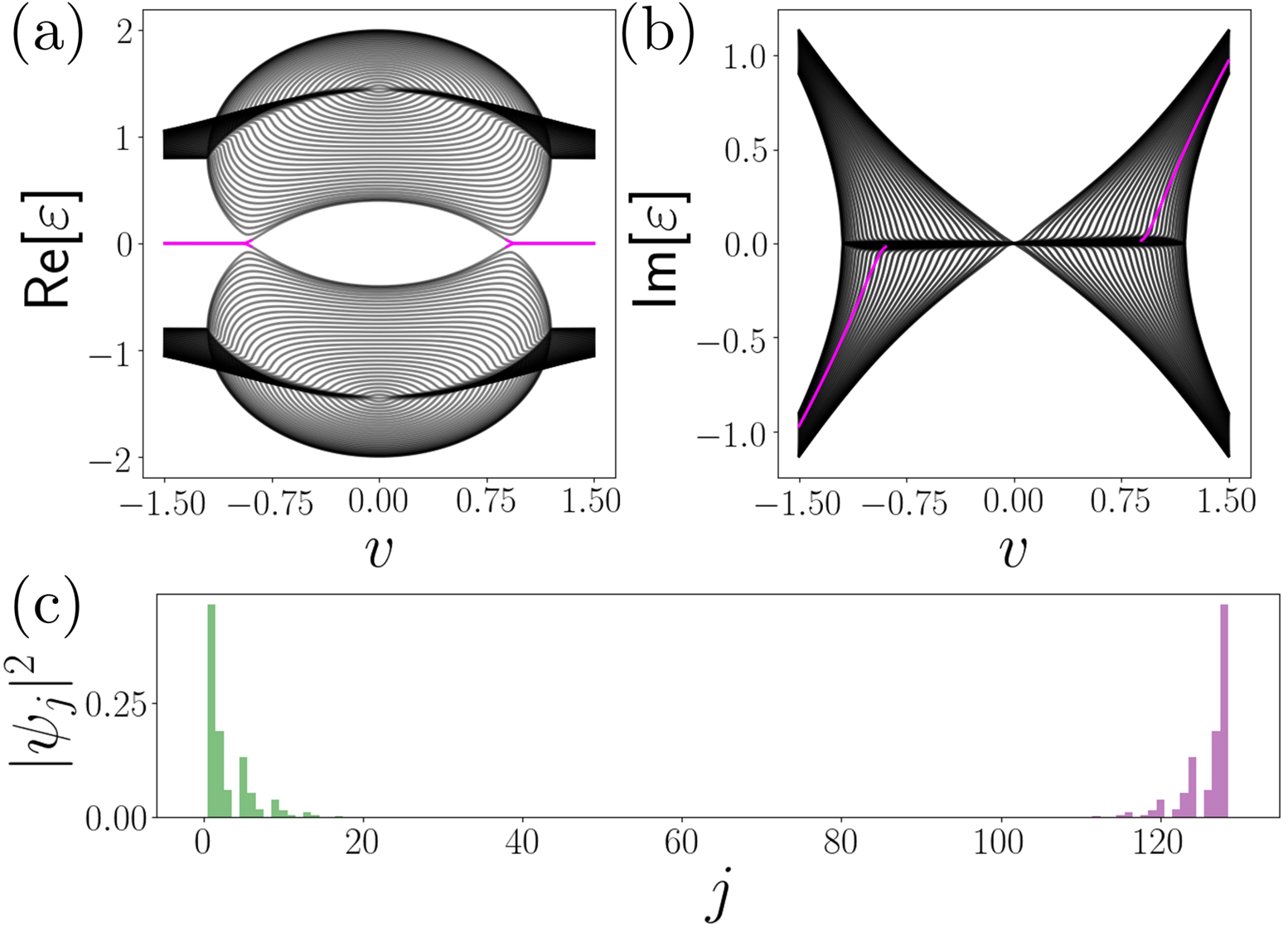}
\end{center}
\caption{(a) The real and (b) imaginary parts of the single-particle spectrum $\varepsilon$ of $H_{\rm hop}+H_{\rm pot}$ with open boundary conditions, i.e., the hopping between the sites $N$ and $1$ is set to zero. 
The results are shown as a function of $v$ for $N=128$ with the parameters $(t,\delta)=(1.0,0.2)$. 
The magenta lines denote twofold degenerate edge states. 
(c) Spatial profiles of the twofold degenerate edge states for $N=128$. 
The parameters are $(t,v,\delta)=(1.0,1.1,0.2)$ [See $\circ$ in Fig.~\ref{BP1}(b)]. The green and purple bars show the right single-particle eigen functions of twofold degenerate edge states.
}
\label{Edge}
\end{figure}

In the absence of interaction ($U=0$), the complex Berry phase for $N=64$ is shown in Fig.~\ref{BP1}(b).
The figure clearly shows that the complex Berry phase is quantized to $\gamma = 0$ or $\pi$.
In this case, phase boundary can be determined analytically.
The Hamiltonian $H_1(0) = H_{\rm hop}+H_{\rm pot}$ in momentum space is represented by a $4 \times 4$ matrix as
\begin{align}
h(k)
= 
\begin{pmatrix}
iv&t+\delta&0&(t-\delta)e^{-ik}
\\
t+\delta&-iv&t-\delta&0
\\
0&t-\delta&-iv&t+\delta
\\
(t-\delta)e^{ik}&0&t+\delta&iv
\end{pmatrix}
,
\end{align}
where $k$ is a wave number.
The single-particle energy levels $\varepsilon_k$ are given by
\begin{align}
\varepsilon_k = \pm \sqrt{2t^2 + 2\delta^2 -v^2 \pm 2(t-\delta)\sqrt{(t+\delta)^2\cos^2\frac{k}{2}-v^2}}.
\end{align}
Therefore, the real spectral gap $\Delta$ of the initial Hamiltonian $H_1(0)$ is evaluated as
$\Delta=2\left|{\rm Re} (t-\delta -\sqrt{(t+\delta)^2-v^2}) \right|$.
The gap closes when $\delta=v^2/4t$, which agrees with the boundary between the phases with $\gamma=0$ and $\pi$, as shown in Fig.~\ref{BP1}(b).
This result indicates that there exists a covalent state between sites $N$ and $1$ in the $\pi$-Berry phase region $\delta<v^2/4t$.
Note that as long as the Berry phase is $\pi$, the covalent state remains there no matter how small the hopping between the sites is.
As seen in Fig. \ref{Edge}, the broken pieces of the covalent state appear as edge states in the decoupled limit \cite{kariyado2014}. 
In Fig.~\ref{Edge}, the magenta lines in~(a) and (b) show the single-particle edge states localized at the edges for the $N=128$ system, whose representative spatial profiles are illustrated in~(c) with the green and purple bars.
They appear in the $\pi$-Berry phase region. 
Actually, these results are consistent with the previous study~\cite{Revision_Hyart2022}, where the appearance of the edge states is confirmed in the $\delta=0$ case.
On the other hand, in the region $\delta>v^2/4t$, the complex Berry phase is quantized to $0$, 
in which case edge states do not appear when the bond between sites $N$ and $1$ is broken.

We now turn to discuss the case where the nearest-neighbor interaction is turned on. 
Figure~\ref{BP1}(c) shows the $U$ dependence of the complex Berry phase.
The other parameters are $t=1.0$, $\delta=0.2$, and $v=1.1$ (see the white circle $\circ$ in Fig.~\ref{BP1}(b)).

The numerical result for $N=16$ suggests that the topological phase transition takes place near $U=0.97$.
The complex Berry phase is quantized as long as it is well-defined, regardless of whether the interaction is attractive or repulsive \footnote{Too strong attractive interaction can lead to ground-state degeneracy, making the complex Berry phases ill-defined}.

\begin{figure}[h]
\begin{center}
\includegraphics[width=7.8cm]{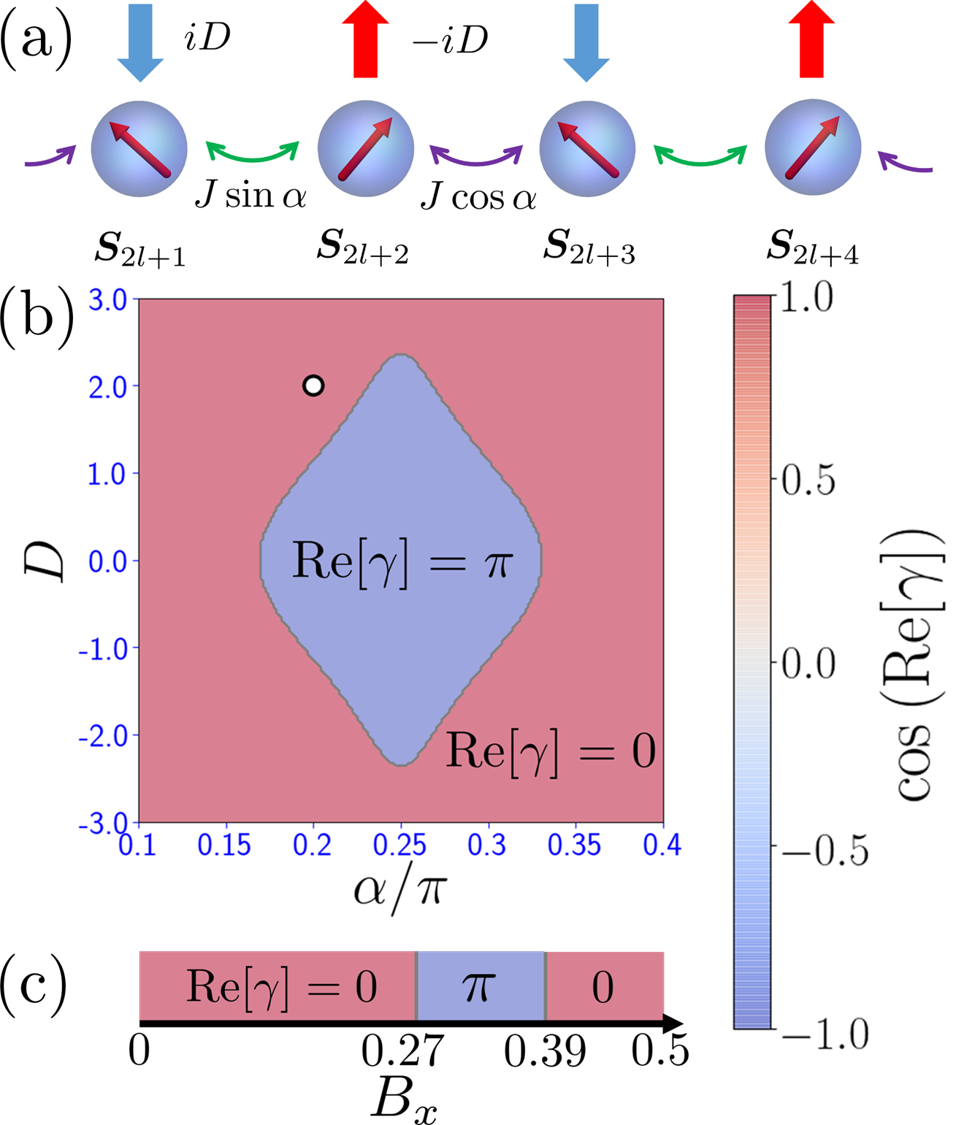}
\end{center}
\caption{(a) Schematic representation of the antiferromagnetic spin chain described by Eq.~(\ref{Ham2}) with $B_x=0$. (b) The real part of the complex Berry phase 
for the model Eq.~(\ref{Twist2}) with $J=1.0$ and $B_x=0$. 
(c) The real part of the complex Berry phase for the model Eq. (\ref{Twist2}) with $(J, \alpha, D)=(1.0, 0.2\pi,2.0)$, $N=10$. 
The corresponding parameter set is indicated by the white circle $\circ$ in (b). 
}
\label{BP2}
\end{figure}

\textit{Non-Hermitian spin-1 chain. ---}
As  a demonstration of the quantized complex Berry phase for Case~\caseb, we study a spin-$1$ antiferromagnetic chain. 
The Hamiltonian of the model is given by
\begin{align}
H_2
&= 
J\sum_{j=1}^N
\left[
{1-(-1)^j\over2}
\sin\alpha +{1+(-1)^j\over2}\cos\alpha
\right]
\bm{S}_{j}\cdot\bm{S}_{j+1}
\nonumber
\\
&
-iD
\sum_{j=1}^N\left[(-1)^j
(S^z_{j})^2
\right]
-iB_x
\sum_{j=1}^N\left[(-1)^j
S^x_{j}
\right],
\label{Ham2}
\end{align}
where $\bm{S}_j=(S_j^x,S_j^y,S_j^z)$ is the spin-$1$ operator at site $j$. We impose 
PBC and assume that $N$ is even. 
The first term is the antiferromagnetic Heisenberg interaction, where $J\sin\alpha$ and $J\cos\alpha$ are alternating couplings with $J>0$, $\alpha \in (0, \pi/2)$. The second and third terms, respectively, describe the imaginary single-ion anisotropy with strength $D \in \mathbb{R}$ and the imaginary transverse magnetic field of strength $B_x \in \mathbb{R}$. 
The schematic picture of the model is shown in Fig.~\ref{BP2}(a).

To define the complex Berry phase, we take a local 
gauge twist, under which the coupling between sites $N$
and $1$ is replaced by the $\theta$-dependent one as
\begin{align}
H_2(\theta) = 
H_2 
+\frac{J}{2}\cos\alpha
\left\{(e^{-i\theta} \!-\!1)
S^+_{N} S^-_{1}
+(e^{i\theta}\!-\!1)S^-_{N} S^+_{1}\right\},
\label{Twist2}
\end{align}
where $S^{\pm}_j= S^x_j \pm i S^y_j$.
The Hamiltonian $H_2(\theta)$ satisfies Eq.~(\ref{inv2}) with the bond-centered inversion $P$: $\bm{S}_j \rightarrow \bm{S}_{N+1-j}$. 
We start with the model without transverse magnetic field, i.e., $B_x=0$. The calculation for an $N=10$ system is shown in Fig.~\ref{BP2}(b). The real part of the complex Berry phase is indeed quantized and provides us with information about the topological phase of the ground state.
In the blue region in Fig.~\ref{BP2}(b), the real part of the Berry phase is $\pi$, which indicates the presence of a spin singlet formed by two fractionalized spin-${1\over2}$'s at site $N$ and site $1$.
In this region, all links are covered by such spin singlets. This region corresponds to a phase equivalent to the Haldane phase in the Hermitian case.
On the other hand, in the red region, the real part of the complex Berry phase is $0$, indicating that this region corresponds to the dimer phase.

\begin{figure}[h]
\begin{center}
\includegraphics[width=8.2cm]{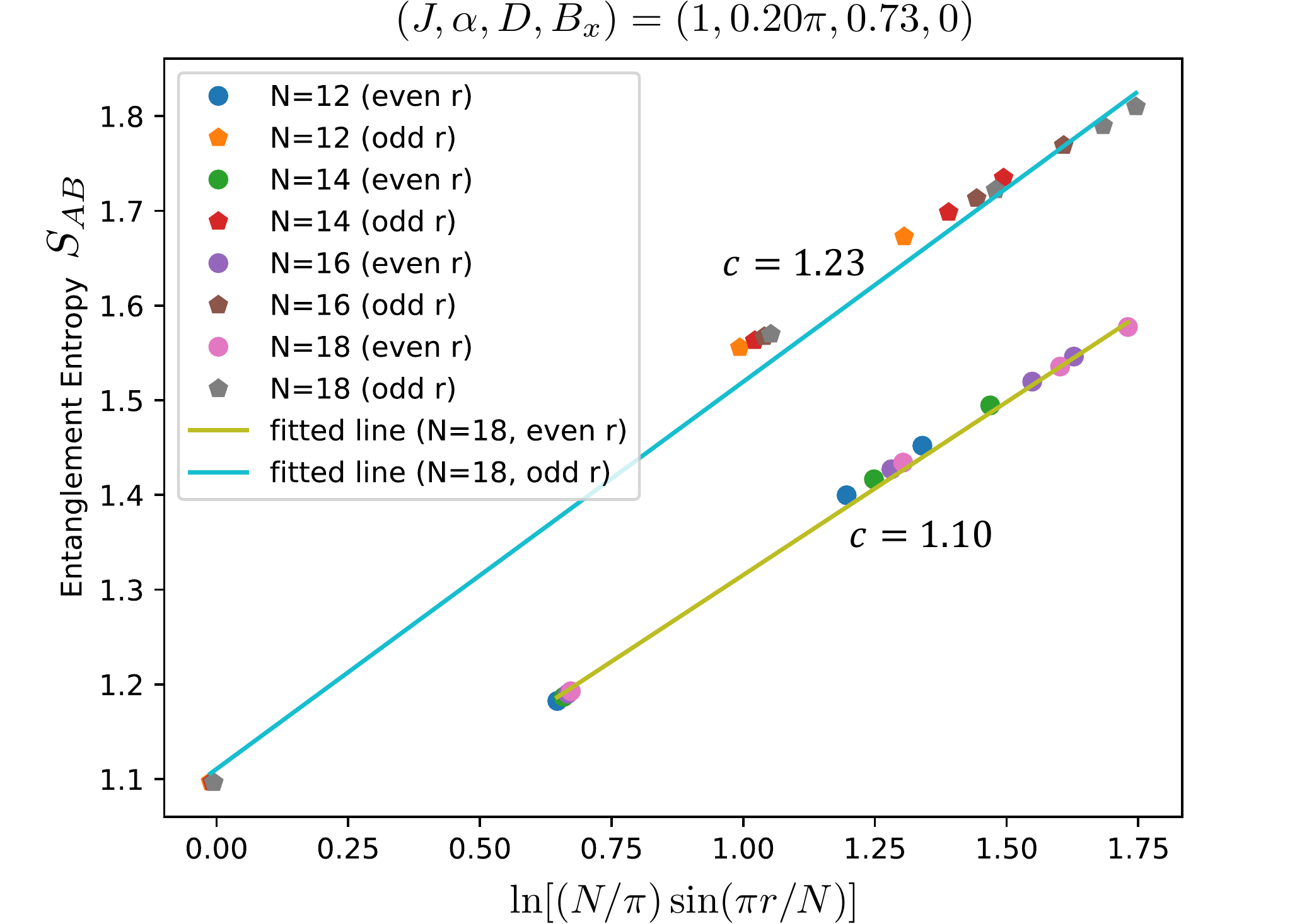}
\end{center}
\caption{Ground-state entanglement entropy $S_{AB}$ obtained by exact diagonalization of non-Hermitian $H_2$. For $N=18$, we find that $S_{AB}(J=1, \alpha = 0.20\pi,D ,B_x = 0)$ has a peak at $D= 0.73$ (not shown), indicating a quantum critical point. The central charge $c$ at the critical point 
is calculated by linear fitting for $N=18$. 
We obtain $c=1.10$ when the subsystem size $r$ is even
while $c=1.23$ when $r$ is odd.
In both cases, $c$ is close to 1.}
\label{CFT}
\end{figure}

In Fig.~\ref{BP2}(b), the Haldane phase and the dimer phase are separated by a critical phase transition. In the Hermitian case ($D=B_x=0$), it is known that the criticality is
described by a conformal field theory with central charge 
$c=1$~\cite{doi:10.1143/JPSJ.63.1277, doi:10.1143/JPSJ.68.2779}.
For a one-dimensional critical system of length $N \gg 1$ with PBC, let $A$ be a segment (subsystem) of length $r$ and $B$ be the rest of the system. The entanglement entropy between $A$ and $B$ is given by $S_{AB} = (c/3)\ln[(N/\pi)\sin(\pi r/N)] + s_1$~\cite{Calabrese2004}, 
where $s_1$ is a non-universal constant. 
To identify $c$ in the non-Hermitian case, we define
$S_{AB} = - \text{tr} \rho_{A} \ln\rho_A$, where the ground-state reduced density matrix 
$\rho_{A}  = \text{tr}_{B} \ket{\Psi(\theta)} \bra{\Psi(\theta)} $ \cite{Hamazaki2019}. 
We calculate $S_{AB}$ for $H_2$ by exact diagonalization and
conclude that $c=1$ in the non-Hermitian critical points, which is consistent with the Hermitian case (see Fig.~\ref{CFT}).

The analysis using quantized complex Berry phases is also valid for systems without $U(1)$ symmetry. 
As an example, we consider the case where $B_x \ne 0$. 
In Fig.~\ref{BP2}(c), we show ${\rm Re}[\gamma]$ as a function of $B_x$ for $N=10$.
The other parameters are $J=1.0$, $\alpha=0.2\pi$, $D=2.0$, denoted by the white circle $\circ$ in Fig.~\ref{BP2}(b). 
Figure ~\ref{BP2}(c) shows that the complex Berry phase is quantized to $\pi$ for $0.27 < B_x < 0.39$. 
We should note, however, that this nontrivial region might be an artifact of finite-size effects and vanish in the thermodynamics limit.
A precise determination of the phase diagram is left for future work.

Finally, we 
make some remarks about the entanglement spectrum (ES). In the Hermitian 
case, it is well-known that the 
Haldane phase is characterized by an even-fold degeneracy of the whole ES~\cite{PhysRevB.81.064439, PhysRevB.85.075125, PhysRevB.86.125441}.
Our preliminary numerical results for $H_2$ show that the even-fold degeneracy persists even in the non-Hermitian Haldane phase region.

\medskip

\textit{Summary and Outlook.---}
We have discussed the quantization of complex Berry phases in non-Hermitian many-body systems. 
We have shown that the complex Berry phase of a non-Hermitian Hamiltonian $H(\theta)$ exhibits $\mathbb{Z}_2$ quantization if there exits a unitary Hermitian operator $P$ that maps $H(\theta)$ to $H(-\theta)$ or $H^\dag (-\theta)$. 
The quantized Berry phase can be used to detect nontrivial entanglement in $H(0)$, thus serving as a useful tool to characterize the topological phases of $H(0)$. We illustrated the power of this approach by applying it to certain one-dimensional 
many-body systems with and without interactions, which are representatives of the above two cases. 
We expect that the quantization of the complex Berry phase is not limited to these two cases. 
It would thus be interesting to generalize the present approach to systems with different symmetries,
for example, point group symmetries in
higher dimensions~\cite{PhysRevX.7.011020, PhysRevB.96.205106, PhysRevResearch.3.023210}.
It would also be intriguing to extend the method to the case of non-diagonalizable Hamiltonians \cite{Read2020}.

\begin{acknowledgments}
We thank Yasuhiro Hatsugai and
Tomonari Mizoguchi for valuable discussions. 
Numerical calculation of the entanglement entropy is implemented by QuSpin~\cite{SciPostPhys.2.1.003, 10.21468/SciPostPhys.7.2.020}.
H.~Y. was supported by Grant-in-Aid for JSPS Research Fellowship for Young Scientists (DC1) No. 20J20715. 
Y.~A. was supported by JSPS KAKENHI Grant Nos. JP17K14352, JP20K14411, and JSPS Grant-in-Aid
for Scientific Research on Innovative Areas “Quantum
Liquid Crystals” (KAKENHI Grant No. JP20H05154 and JP22H04469). 
H.~K. was supported by JSPS Grant-in-Aid for Scientific Research on Innovative Areas No. JP20H04630, JSPS Grant-in-Aid for Scientific Research No. JP18K03445, Grant-in-Aid for Transformative Research Areas (A) “Extreme Universe” No. JP21H05191[D02], and the Inamori Foundation.
\end{acknowledgments}

\bibliographystyle{apsrev4-1}
\bibliography{refs}

\clearpage


\end{document}